\documentclass[submission,copyright,creativecommons]{eptcs}
\providecommand{\event}{AREA2022} 
\usepackage{breakurl}             
\usepackage{underscore}           
\usepackage{graphicx} 
\usepackage{listings}
\usepackage{color}
\usepackage{xspace}

\usepackage{comment}

\title{RV4JaCa -- Runtime Verification for \\Multi-Agent Systems} 

\author{Debora C. Engelmann
\institute{School of Technology -- PUCRS, Brazil}
\institute{DIBRIS -- University of Genoa, Italy}
\email{debora.engelmann@edu.pucrs.br}
\and
Angelo Ferrando
\institute{DIBRIS -- University of Genoa, Italy}
\email{angelo.ferrando@unige.it}
\and
Alison R. Panisson
\institute{Department of Computing -- UFSC, Brazil}
\email{alison.panisson@ufsc.br}
\and
Davide Ancona
\institute{DIBRIS -- University of Genoa, Italy}
\email{davide.ancona@unige.it}
\and
Rafael H. Bordini
\institute{School of Technology -- PUCRS, Brazil}
\email{rafael.bordini@pucrs.br}
\and
Viviana Mascardi
\institute{DIBRIS -- University of Genoa, Italy}
\email{viviana.mascardi@unige.it}
}
\def\titlerunning{RV4JaCa -- RV for MAS} 
\def\authorrunning{\xspace D.C. Engelmann, A. Ferrando, A.R. Panisson,\\ \xspace D. Ancona, R.H. Bordini \& V. Mascardi}
\begin{document}

\lstset{
	morekeywords={not,none,any,empty,matches,all,let,if,else,silent,id,topics,node,log,nodes,monitor,monitors,with},
	keywordstyle=\color{blue},
	morestring=[b]',
	stringstyle=\color{red},
	morecomment=[l]{//},
	commentstyle=\color{olive},
	mathescape=true,
	basicstyle=\scriptsize\ttfamily,
	captionpos=b,
	tabsize=4,
	breaklines,
	breakatwhitespace,
	showstringspaces=false,
	keepspaces,
	numbers=left,
	frame=single
}

\maketitle

\begin{abstract}
This paper presents a Runtime Verification (RV) approach for Multi-Agent Systems (MAS) using the JaCaMo framework. Our objective is to bring a layer of security to the MAS. This layer is capable of controlling events during the execution of the system without needing a specific implementation in the behaviour of each agent to recognise the events.
MAS have been used in the context of hybrid intelligence. This use requires communication between software agents and human beings. In some cases, communication takes place via natural language dialogues. However, this kind of communication brings us to a concern related to controlling the flow of dialogue so that agents can prevent any change in the topic of discussion that could impair their reasoning.
We demonstrate the implementation of a monitor that aims to control this dialogue flow in a MAS that communicates with the user through natural language to aid decision-making in hospital bed allocation.

\end{abstract}

\section{Introduction}

A characteristic identified as essential in Artificial Intelligence (AI) is explainability, as it provides users with the necessary inputs to make it possible to understand the system's behaviour, and inspire confidence in the final outcome.
Explainability becomes an essential feature in Multi-Agent Systems (MAS), as it is one of the most powerful paradigms for implementing complex distributed systems powered by artificial intelligence techniques.
MAS are built upon core concepts such as distribution, reactivity, and individual rationality. Agents have been widely studied, and an extensive range of tools have been developed, such as agent-oriented programming languages and methodologies~\cite{Bordini-2009-MAPLTA}. Thus, practical applications of multi-agent technologies have become a reality to solve complex and distributed problems~\cite{schmidt2016ontology}.
In addition, it also allows the execution of various tasks and makes it possible the integration with various technologies.

Even though MAS solutions can be a natural choice for developing complex and distributed systems, like any other software development technique they are prone to errors and bugs (whether at the implementation or description level). Standard techniques such as testing and debugging can be deployed to tackle this problem. However, in case of MAS, the process of testing~\cite{DBLP:journals/aamas/Winikoff17}, debugging~\cite{DBLP:conf/atal/Winikoff17}, and verifying~\cite{DBLP:journals/ase/DennisFWB12} such systems can be quite complex.
For this and other reasons, more lightweight approaches to guarantee the correct execution of the system are valuable.
One technique that can be applied in such cases is Runtime Verification (RV)~\cite{DBLP:series/lncs/BartocciFFR18,DBLP:journals/jlp/LeuckerS09}. Differently from other verification techniques, RV is lightweight because it only concerns the analysis of the runtime execution of the system under analysis; which makes RV very similar to testing. However, rather than testing, RV is based on a specification formalism, as it happens in formal verification, to express the properties to be checked against the system's behaviour, and is particularly suitable to monitor
control-oriented properties \cite{AhrendtCPS17}.

In this paper, we present RV4JaCa, an approach to perform RV of multi-agent systems developed using the JaCaMo framework~\cite{boissier2013multi}. RV4JaCa is obtained by extending MASs implemented in JaCaMo with a monitoring feature. In a nutshell, RV4JaCa handles all the engineering pipeline to introduce RV in JaCaMo, and to extract and check runs of the MAS against formal properties. In particular, RV4JaCa enables RV of agents' interactions (i.e., messages). As a proof of concept, we demonstrate a case study in a hospital bed allocation domain where RV is used to verify two different agent interaction protocols.

The paper is structured as follows. Section~\ref{sec:background} reports the background needed to fully understand the contribution; here, the JaCaMo framework and the notion of RV are introduced.  Section~\ref{sec:rv4jaca} presents RV4JaCa and its application in a bed allocation case study. Section~\ref{sec:related} positions the contribution w.r.t. the state of the art. Finally, Section~\ref{sec:conclusions} summarises the contribution's results, and points out future developments.

\section{Background}
\label{sec:background}

\subsection{Multi-agent systems}

Multi-agent systems (MAS) are systems composed of multiple agents.
They seem to be a natural metaphor for building and understanding a wide range of artificial social systems and can be applied in several different domains~\cite{WooldridgeAnIntroduction:2002}.
There are two interlocking strands of work in multi-agent systems: the one that concerns individual agents and the one that concerns itself with the collections of these agents.
In practice, agents rarely act alone, they usually inhabit an environment that contains other agents. Each agent can control, or partially control, parts of the environment, which is called its ``sphere of influence''.
It may happen that these spheres of influence overlap, hence (parts of) the environment may be controlled jointly by more than one agent. In this case, to achieve the desired result, an agent must also consider how other agents may act. These agents will have some knowledge, possibly incomplete, about the other agents~\cite{bordini2007programming}.

Wooldridge~\cite{WooldridgeAnIntroduction:2002} believes that to be able to understand a multi-agent domain it is essential to understand the type of interaction that occurs between agents.
For intelligent autonomous agents, the ability to reach agreements is extremely important, and for this, negotiation and argumentation skills are often necessary.
In~\cite{FerberAtAllFrom:2003}, the authors mention two types of studies in multi-agent systems; the first one is Agent-Centred Multi-Agent Systems (ACMAS), which study, at the level of an agent, states and the relationship between those states and their general behaviour and are projected in terms of the agent's states of mind.
The second one is Organisation-Centred Multi-Agent Systems (OCMAS), which are systems whose foundations reside in the concepts of organisations, groups, communities, roles, and functions, among others. An OCMAS is not considered in terms of mental states but in capacities and constraints, which are considered organisational concepts, as well as functions, tasks, groups, and interaction protocols.

In a multi-agent system, the organisation is the collection of roles, relationships, and authority structures that govern agents' behaviour.
Every multi-agent system has some form of organisation, even if it is implicit and informal.
Organisations guide the mode of interaction between agents, which may influence data flows, resource allocation, authority relationships, and various other features of the system~\cite{HorlingAndLesserASurvey:2004}.

In~\cite{Anjomshoae-2019-EAandRRfromaSLR}, the authors argue that systems which heavily adopt AI techniques are increasingly available, and making them explainable is a priority.
Explainable Artificial Intelligence (XAI) is a research field that aims ``to make AI systems results more understandable to humans''~\cite{adadi2018peeking}. These results must be clear (in non-technical terms) and provide justifications about decisions made~\cite{donadello2020explaining}.
In order to achieve a satisfactory level of explainability in multi-agent systems, much communication between agents and humans needs to be performed. However, this can generate an additional point of failure in the execution of the system since if, for example, the predefined communication protocol is not followed, or even if the human decides to change the conversation topic unexpectedly, this can lead to unexpected and probably inappropriate behaviour of the system.

\subsection{JaCaMo framework}

\begin{figure}[h]
    \centering
    \includegraphics[width=0.80\textwidth]{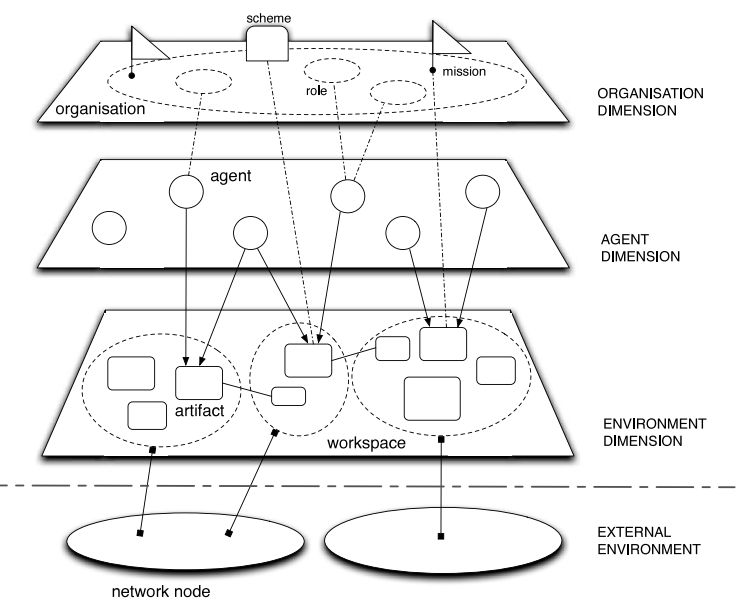}
    \caption{Overview of the three dimensions of JaCaMo \cite{boissier2013multi}}
    \label{figJACA}
\end{figure}

JaCaMo is a framework that allows for multi-agent oriented programming. This framework consists of the integration of three previously existing platforms: Jason -- for programming autonomous agents, CArtAgO -- for programming environmental artifacts, and Moise -- for programming multi-agent organisations~\cite{boissier2013multi}. A multi-agent system programmed in JaCaMo has Jason agents that are organised and follow roles according to Moise's hierarchical structure. These agents work in environments based on distributed artifacts programmed using CArtAgO. Figure~\ref{figJACA} shows an overview of the three dimensions of JaCaMo.

Jason (\texttt{Agent dimension}) is an agent-oriented programming language that is an interpreter for an extended version of the AgentSpeak~\cite{bordini2007programming} language. Agents programmed in Jason use the Belief-Desired-Intention (BDI) model~\cite{rao1995bdi}. The main idea of this approach is to model the process of deciding which action to take to achieve certain objectives~\cite{rao1995bdi}. Moise~\cite{hubner2007developing,boissier2013multi} is related to the \texttt{organisation dimension}, where agents can be part of groups and follow specific roles. Also, with Moise, \emph{schemes} are defined, that is, the structure of organisation goals is decomposed into sub-goals and grouped into missions. An organisation is specified in an XML (Extensible Markup Language) file.
CArtAgO~\cite{ricci2009environment}, the \texttt{environment dimension}, is used to simulate an environment or interface with a real one; this is where \emph{artifacts} are defined. These artifacts define the environment's structure and behaviour, representing all resources that agents need. Agents can discover, create, and use artifacts at runtime~\cite{boissier2013multi}. Artifacts are programmed in Java. Combining these dimensions provides us with a complete framework for developing multi-agent systems through agents, organisations, and environments.

\subsection{Runtime verification}

Runtime Verification (RV)~\cite{bartocci2018introduction} is a kind of formal verification technique that focuses on checking the behaviour of software/hardware systems. With respect to other formal verification techniques, such as Model Checking~\cite{clarke1997model} and Theorem Provers~\cite{DBLP:books/lib/Loveland78}, RV is considered more dynamic and lightweight. This is mainly due to its being completely focused on checking how the system behaves, while the latter is currently running. This is important from a complexity perspective. RV does not need to simulate the system in order to check all possible execution scenarios; but, it only analyses what the system produces (i.e., everything that can be observed of the system). This is usually obtained through monitors, which are automatically generated from the specifications of the properties to be checked, and are nothing more than validation engines  which, given a trace of events generated by the system execution, conclude the satisfaction (resp. violation) of the corresponding properties. In turn, a formal property is the formal representation of how we expect the system should behave. The monitor's job is to verify at runtime whether such a property holds.

Since monitors are usually deployed together with the system under analysis, they are well suited for checking properties that require to be continuously monitored. This is especially true in safety-critical scenarios, where a system's fault can cause injuries, loss of money and even deaths. A key example is autonomous and robotic systems, where \emph{reliability} is vital~\cite{DBLP:journals/aamas/FisherMRSWY21}, and the addition of monitors ensuring a correct behaviour is a valuable feature.

In the scenario envisaged in this contribution, we aim to use RV as a safety net for message exchange in JaCaMo. As pointed out elsewhere, the protocols involved in the communication amongst agents and human beings can be very complex, and hard to track. Moreover, agents are usually particularly focused on the reasoning and reactive aspects, while the consistency of the protocols is given for granted. However, above all in case of human being in the loop, such assumption cannot be made. RV is a suitable candidate to keep track of the protocols, to check whether the current agents' enactment is consistent (or not) with the expected protocol. Such consistency checking is extremely important in safety-critical scenarios, as in the healthcare domain, where a protocol violation can be costly.

\subsection{Runtime Monitoring Language}

Runtime Monitoring Language\footnote{\url{https://rmlatdibris.github.io/}} (RML \cite{AnconaFFM21}) is a Domain-Specific Language (DSL) for specifying highly expressive properties in RV (such as non context-free ones). We use RML in this paper for its support of parametric specifications and its native use for defining interaction protocols. In fact, the low-level language on which RML is based upon was born for specifying communication protocols~\cite{DBLP:conf/birthday/AnconaFM16,DBLP:conf/atal/AnconaFM17}.

Since RML is just a means for our purposes, indeed other formalisms can be as easily integrated into RV4JaCa, we only provide a simplified and abstracted view of its syntax and semantics. However, the complete presentation can be found in~\cite{AnconaFFM21}.

In RML, a property is expressed as a tuple $\langle t,\mathit{ETs} \rangle$, with $t$ a term and $\mathit{ETs}=\{ET_1,\ldots,ET_n\}$ a set of event types. An event type $ET$ is represented as a set of pairs $\{k_1:v_1,\ldots,k_n:v_n\}$, where each pair identifies a specific piece of information ($k_i$) and its value ($v_i$). An event $Ev$ is denoted as a set of pairs $\{k_1':v_1',\ldots,k_m':v_m'\}$. Given an event type $ET$, an event $Ev$ matches $ET$ if $ET \subseteq Ev$, which means $\forall (k_i:v_i) \in ET \cdot \exists (k_j:v_j) \in Ev \cdot k_i = k_j \land v_i = v_j$. In other words, an event type $ET$ specifies the requirements that an event $Ev$ has to satisfy to be considered valid.

An RML term $t$, with $t_1$, $t_2$ and $t'$ as other RML terms, can be:
\begin{itemize}
    \item $ET$, denoting a set of singleton traces containing the events $Ev$ s.t. $ET \subseteq Ev$;
    \item $t_1 \;\; t_2$, denoting the sequential composition of two sets of traces;
    \item $t_1 \; | \; t_2$, denoting the unordered composition of two sets of traces (also called shuffle or interleaving);
    \item $t_1 \land t_2$, denoting the intersection of two sets of traces;
    \item $t_1 \lor t_2$, denoting the union of two sets of traces;
    \item $\{ let \; x;\; t' \}$, denoting the set of traces $t'$ where the variable $x$ can be used (i.e., the variable $x$ can appear in event types in $t'$, and can be unified with values).
    \item $t' *$, denoting the set of chains of concatenations of traces in $t'$
\end{itemize}


Event types can contain variables. For example, $ET(ag1, ag2)=\{sender:ag1, receiver:ag2\}$, where we do not force any specific value for the sender (resp., receiver) of a message (in this case the events of interest would be messages). 
This event type matches all events containing sender and receiver. When an event matches an event type with variables, such as in this case, the variables get the values from the event. For instance, if the event observed would be $Ev=\{sender:``Alice", receiver:``Bob"\}$, it would match $ET$ by unifying its variables as follows: $ag1=``Alice"$, and $ag2=``Bob"$. This aspect is important because, as we are going to show in the bed allocation domain, we can use variables in RML terms to enforce a specific order of messages. For instance, in this very high-level example, we could say that when a message from $ag1$ to $ag2$ is observed, the only possible consequent message can be a message from $ag2$ to $ag1$. Since the first event has unified the two variables, the second event will have to be a message from $Bob$ to $Alice$ (otherwise this would be considered a violation). Naturally, this is only the intuition behind it, but it should help to grasp the expressiveness of RML and how variables can be exploited at the protocol level to enforce specific orders amongst the messages.

\section{Engineering runtime verification for multi-agent systems}
\label{sec:rv4jaca}

\begin{figure}
    \centering
    \includegraphics[scale=0.6]{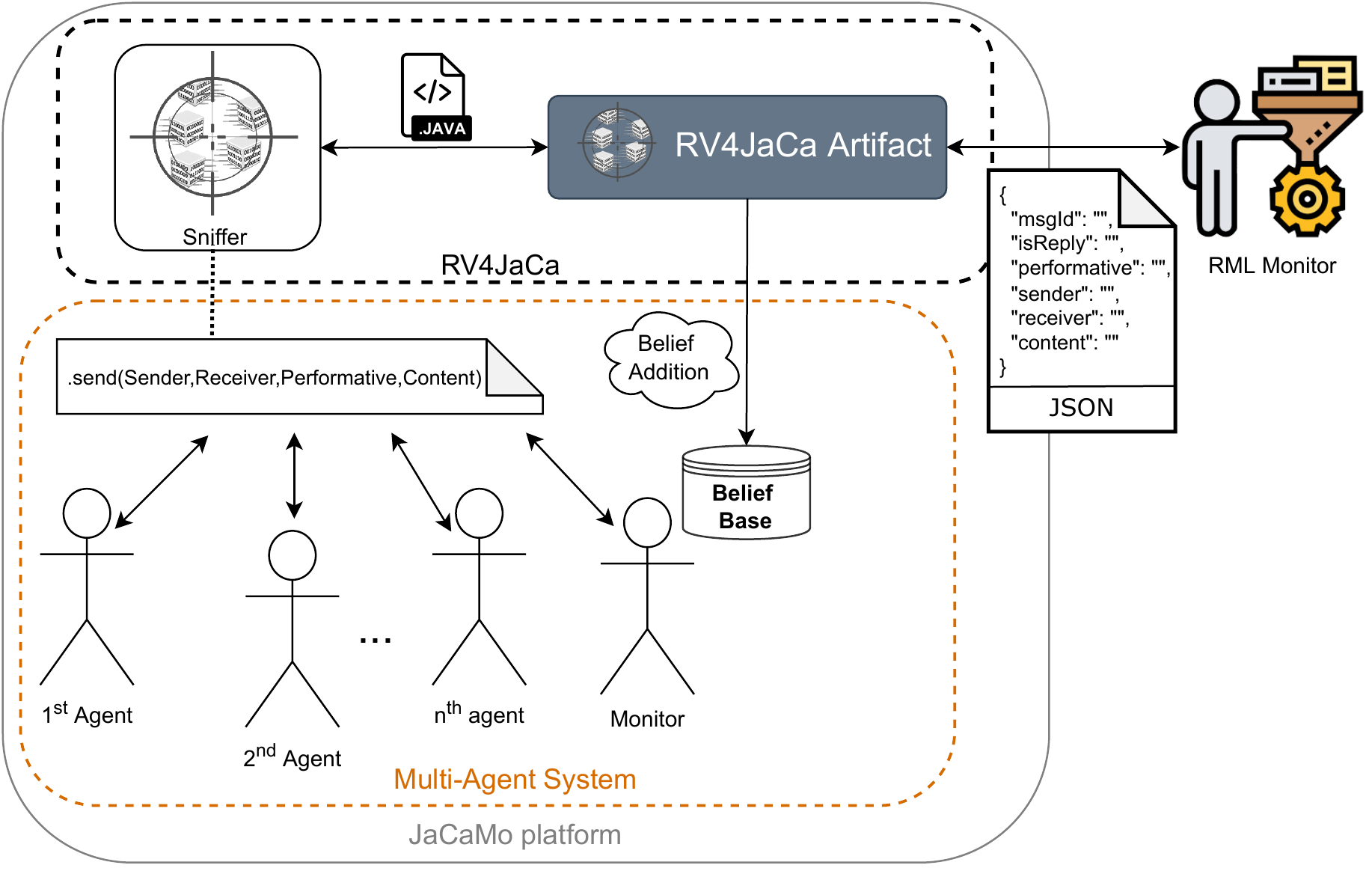}
    \vspace{-5pt}
    \caption{Approach overview for RV in MAS.}
    \label{fig:RV4JaCaArchitecture}
\end{figure}

Our approach named RV4JaCa\footnote{The source code is available at \url{https://github.com/DeboraEngelmann/RV4JaCa}} allows the runtime verification of multi-agent systems based on the JaCaMo platform. In Figure~\ref{fig:RV4JaCaArchitecture}, we present an overview of the entire approach.
RV4JaCa is composed of:
(i) a \texttt{Sniffer} class, developed in Java, responsible for observing all communication between agents in the MAS;
(ii) a CArtAgO artifact named \texttt{RV4JaCa Artifact} responsible for analysing the messages observed by the \texttt{Sniffer}, transforming them into a JSON (JavaScript Object Notation) object and sending it as a REST (Representational State Transfer) request to the \texttt{RML Monitor}. Note that RV4JaCa is not in any way limited to a specific kind of monitor; we used RML simply because it was a most suitable candidate for specifying the protocols of our interest. Nonetheless, a different monitor could be as easily integrated as the RML one. In addition, when the \texttt{RV4JaCa Artifact} receives the response of the request made to the \texttt{RML Monitor} saying that there was a violation, it can add a belief in the \texttt{Monitor} agent belief base;
(iii) the \texttt{RML Monitor} responsible for analysing the events sent by \texttt{RV4JaCa Artifact} and verifying the satisfaction or violation of a formal property of interest; and
(iv) a \texttt{Monitor} agent, which can be added to the system if it is necessary to interfere with agents' behaviour at runtime. In this case, if there is a violation, the \texttt{RV4JaCa Artifact} adds a belief to the \texttt{Monitor}'s belief base. When the agent perceives this addition, it can react by sending a message to the interested agents warning about the violation. This may trigger some consequent recovery mechanism, which usually is fully domain dependent. On the other hand, the \texttt{Monitor} agent can also perform different activities depending on the needs of the system.
To clarify how our approach works, we developed a case study in the field of bed allocation.

\subsection{Bed allocation case study}

\begin{figure}
    \centering
    \includegraphics[scale=0.7]{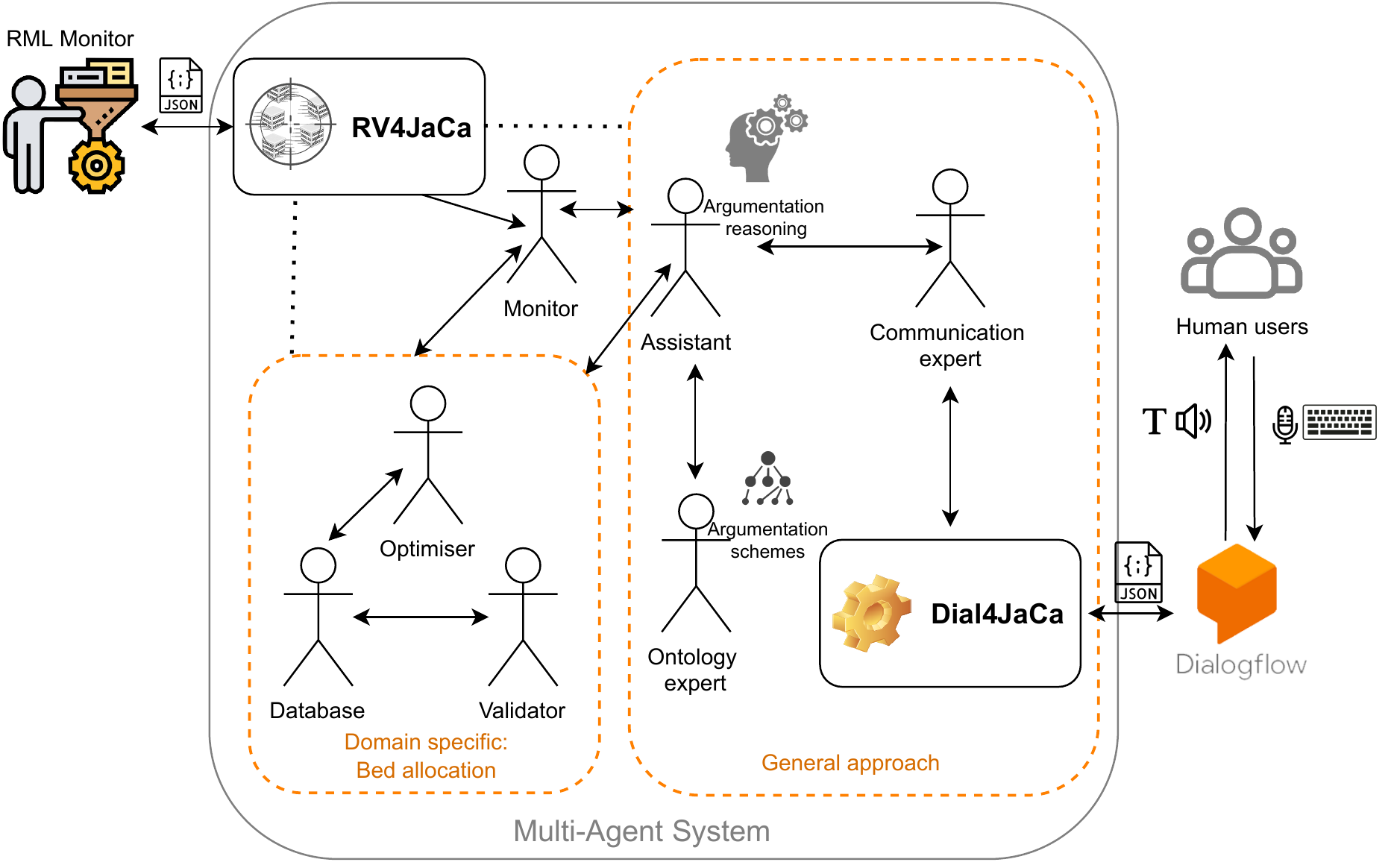}
    \vspace{-5pt}
    \caption{RV4JaCa in a bed allocation domain}
    \label{fig:RV4JaCaBedAllocation}
\end{figure}

We are working on a framework that supports the development of multi-agent applications to assist humans in decision making. One of the domains for which we are building an instance of this framework is hospital bed allocation.
As it is shown in Figure~\ref{fig:RV4JaCaBedAllocation}, to provide interface with natural language processing platforms, such as Dialogflow\footnote{https://cloud.google.com/dialogflow/es/docs}, our framework relies on the use of Dial4JaCa~\cite{engelmann2021dial4jaca}.
The \texttt{Human user} can interact with the chatbot through text or voice. Dialogflow classifies the interaction intents and sends it to Dial4JaCa, making the request available to the \texttt{Communication expert} agent assigned to that specific user.
One or more \texttt{Communication expert} agents can be instantiated, each one responsible for representing one particular \texttt{Human user}. It uses natural language templates~\cite{panisson2021engineering} to translates the \texttt{Assistant} responses (the result of the MAS reasoning) into natural language and send them to its corresponding \texttt{Human user}.
The \texttt{Assistant} agent performs argumentation reasoning~\cite{Panisson-2014-AnAforABRUDLinMAPL} and is responsible for communicating with other agents in search of information.
Several \texttt{Ontology expert} agents can be instantiated, allowing the MAS to consult different ontologies simultaneously given that each of these agents are able to interface with a specific ontology through a CArtAgO artifact.
Such agents can also perform ontological reasoning.
Dial4JaCa together with these three types of agents make up our \texttt{General approach}.

We can add domain-specific agents to the system to address the specificity of different application domains. For example, in the instance shown in Figure~\ref{fig:RV4JaCaBedAllocation}, we added specific agents for the bed allocation domain\footnote{This scenario is detailed in~\cite{engelmann2021conversational}}.
Among those domain-specific agents, the \texttt{Validator} agent is responsible for validating bed allocation plans made by the user (via our system interface) using a PDDL (Planning Domain Definition Language) plan validator; the \texttt{Optimiser} agent is responsible for making suggestions for optimised allocations using the GLPSol solver of GLPK\footnote{http://winglpk.sourceforge.net/} (GNU Linear Programming Kit), which is a free open source software for solving linear programming problems; and the \texttt{Database} agent is responsible for querying and updating the bed allocation system database.

RV4JaCa has been added to that MAS for collecting information about all messages exchanged between agents and sending them through a REST request to the RML monitor (where the properties that need to be checked are defined).
After processing a received message, the monitor returns a result that states whether the message sent from one agent to the other violates any of the properties being checked by it. If a property is violated, RV4JaCa sends the information to the \texttt{Monitor} agent so it can add that information to its belief base and warns the agents involved in the exchange of messages that there has been a violation. This makes it possible for our agents to take action to recover from the failure that the breach caused.

The ability to monitor the messages exchanged between the agents in this case study is used for two different purposes.
The first one is, according to the performative and content of each move, to verify whether the agents are following the predefined communication protocol. This aspect may be crucial for safety-critical and privacy-preserving aspects. For instance, in a healthcare domain, such as bed allocation, the agents might be expected to follow some specific medical guidelines for the communication of personal information (even amongst themselves). Moreover, when in presence of multiple agents, each one having its own goals, it is common to specify the ideal expected outcome at a more abstract level, where is more natural to reason upon.

The second one is to check if a human participant changed the topic of the conversation without the proper conclusion of the previous topic. When we add humans to the agent-to-agent communication loop, the developer has limited control over the human's interactions in the dialogue. For example, in some cases, when the MAS is performing some specific task, it is important to be able to finish it before starting a new one. But when the completion of this task depends on some human interaction, we have no guarantee that the human will complete the necessary exchange. Naturally, it would be possible to do this check within each agent. However, when we have agents specialised in specific tasks, it is preferable to have all the agent plans related to the specific topic rather that other concerns such as verification, which significantly facilitates implementation and code maintenance.

Below, we report two example properties, written in RML, that have been checked for the bed allocation case study using RV4JaCa.

\subsection{RML properties for the bed allocation domain}

The first property, which is presented in Listing~\ref{list:result}, concerns checking that the user does not change the topic before completing the one currently processed by the agents. In particular, the reported property cares about the \lstinline{'getValidationResult'} topic. Such a topic relates to the user asking the assistant agent to validate a suggested bed allocation (the corresponding event type is expressed in lines 1-5). When this message is received by the assistant, the protocol goes on, causing a sequence of messages exchanged amongst the assistant, the optimiser, and the validator agents. Note that this part is not reported because it is not of interest for checking the property. After this step, the assistant agent sends back an answer to the user (the event type is in lines 6-11). If this answer is not empty (i.e., the event contains fields \lstinline{arg1} and \lstinline{arg2}), then the user is expected to conclude the communication with a certain content (listed in lines 12-17). For instance, the user could reply with an additional message containing \lstinline{'allocValPatients'}, meaning that the user is fine with the result of the validation, and he/she wants to allocate the corresponding patient to the proposed bed. Naturally, the user might decline the allocation, in that scenario the message would have content \lstinline{'dontAllocValPatients'}. Similar reasoning goes for the other possible options listed in lines 12-17.

Once the events corresponding to the messages previously mentioned have been specified (lines 1-27), the actual property can be expressed following the RML syntax (lines 28-31). In more detail, we may find in line 28 the definition of the main term denoting the property to check (which in RML is always called \lstinline{Main}). 
In this scenario, the principal term corresponds to a sequence of subterms, named \lstinline{Question}. Such term is defined in line 29, and starts with a question (as defined in lines 1-5). This means that, to comply with the protocol, the first event has to be a message containing a \lstinline{'question'} from the \lstinline{'operator'} to the \lstinline{'assistant'} (in this case regarding the validation of a bed allocation). After that, the property goes on with the \lstinline{Answer} term (line 30). Inside it, we find a disjunction between two possible alternatives in the protocol. On the left, we may observe an \lstinline{answer_with_constraint} event, which means, according to lines 6-11, that the \lstinline{'assistant'} replied to the \lstinline{'operator'} with a validated result that the latter has to decide upon. On the right, we may observe an event corresponding to any other answer, which in this specific case denotes the case where the result is empty, meaning that no result is available to be sent to the \lstinline{'operator'}. In the latter case (the right branch), the property ends this cycle, since the communication between the two agents is concluded and new messages concerning new topics can be exchanged in the future. Instead, in the former case (the left branch), the current cycle is not ended, because the \lstinline{'assistant'} is still waiting for an answer from the \lstinline{'operator'} regarding the result sent. This last aspect is handled in the term in line 31, where no \lstinline{'question'} is admissible from the \lstinline{'operator'}, only one from those listed in lines 12-17. Upon receiving an event matching one such listed event types, the cycle of the property ends, and as for the right branch, the protocol can move on.

Now, before presenting another property of interest that we have analysed through RV4JaCa, it is important to explicit how a property can be violated. As we mentioned before, we presented which are the events that in certain points of the property are accepted, and why. An RML property is violated whenever given the current term denoting the current state of the property, and a new event, the property does not accept such event. For instance, in the property presented in Listing~\ref{list:result}, an event which is different from an answer, after having observed a question, is not acceptable. This can be seen in line 30, where after consuming an event denoting a question, the only possible following events can be an answer requiring additional info (left branch), or a general answer (right branch). Thus, if the observed event is neither of the two, the term is stuck and cannot move on. In RML this translates into a violation of the property, which is then reported back to the monitor agent that in turn will trigger all mechanisms for the agents involved to properly react.

\begin{lstlisting} [xleftmargin=7.0ex,xrightmargin=7.0ex,float=t,caption={The RML specification for checking that no change of topic is observed after a validation result has been requested by the user.},captionpos=b, label={list:result}]
question matches {
    performative:'question',
    sender:'operator', receiver:'assistant',
    content:{name:'getValidationResult'}
};
answer_with_constraint matches
{
    performative:'assert',
    sender:'assistant', receiver:'operator',
    content:{name:'answer', name:'result', arg1:_, arg2:_}
};
constrained_question matches
    {performative:'question', sender:'operator', receiver:'assistant', content:{name:'allocValPatients'}} |
    {performative:'question', sender:'operator', receiver:'assistant', content:{name:'getOptimisedAllocation'}} |
    {performative:'question', sender:'operator', receiver:'assistant', content:{name:'dontAllocValPatients'}} |
    {performative:'question', sender:'operator', receiver:'assistant', content:{name:'allocValidValPatients'}} |
    {performative:'question', sender:'operator', receiver:'assistant', content:{name:'allocValPatients'}};
a_question matches
{
    performative:'question',
    sender:'operator', receiver:'assistant'
};
an_answer matches
{
    performative:'assert',
    sender:'assistant', receiver:'operator'
};
Main = Question*;
Question = (question Answer);
Answer = (answer_with_constraint ConstrainedQuestion) \/ (an_answer);
ConstrainedQuestion = constrained_question Answer;
\end{lstlisting}

The second RML property we tested in the bed allocation domain is reported in Listing~\ref{list:question-answer}. Differently from the property reported in Listing~\ref{list:result}, here we do not check the consistency amongst topics; instead, we care about checking that an agent always replies to a question, before posing a new one. As before, the first part of Listing~\ref{list:question-answer} concerns the definition of which events are of interest for the property (lines 1-10). In this specific case, we have questions (lines 1-5), and answers (lines 6-10). Note that, differently from the previous RML property, here we exploit parameters inside the specification. In fact, the event types reported in lines 1-10 are all parametric w.r.t. the agents involved in the communication. This means that such event types do not focus on specific messages exchanged between predefined agents, as in the previous case, but are kept free (through RML parameters, we have late binding on the agents involved in the interaction). This makes the definition of the RML property in line 11 highly parametric, avoiding the need to update the property for each new agent added to the system.
The property is defined in line 11, through the standard \lstinline{Main} term in RML. Since the property is parametric, it starts with the \lstinline{let} operator which defines the variables used in the term. In this case, the variables used are \lstinline{ag1} and \lstinline{ag2} (naturally any other name would have sufficed). After that, the property goes on expecting a question, followed by a corresponding answer. Here, note that in the first event (i.e., the question), the variables are bound to the agents involved in the communication, while in the second event (i.e., the answer), such variables are ground to the previously initialised values. In this way, a question is free to be sent by any possible agent \lstinline{ag1}, to any possible agent \lstinline{ag2} in the system (where \lstinline{ag1} and \lstinline{ag2} are bound to the observed agents involved in the communication); instead, an answer is constrained to be sent by agent \lstinline{ag2} to agent \lstinline{ag1} (with both variables already bound to the respective values through the previously observed question).

As before, also with this property we can ponder on which events can cause a violation. In particular, the property expressed in Listing~\ref{list:question-answer} is violated when after a question between two agents (\lstinline{ag1}$\rightarrow$\lstinline{ag2}), the following event is not the corresponding answer (\lstinline{ag2}$\rightarrow$\lstinline{ag1}), but another message (for instance another question).

\begin{lstlisting} [xleftmargin=7.0ex,xrightmargin=7.0ex,float=t,caption={The RML specification for checking that an agent always replies before sending messages about something else.},captionpos=b, label={list:question-answer}]
question(ag1, ag2) matches
{
    performative:'question',
    sender:ag1, receiver:ag2
};
answer(ag1, ag2) matches
{
    performative:'assert',
    sender:ag1, receiver:ag2
};
Main = {let ag1, ag2; question(ag1, ag2) answer(ag2, ag1)}*;
\end{lstlisting}

\section{Related Work}
\label{sec:related}

In past years, some work has focused on formal verification from a more dynamic viewpoint.
In~\cite{DBLP:conf/atal/AnconaFM17}, the authors presented a framework to verify Agent Interaction Protocols (AIP) at runtime. The formalism used in this work allows the introduction of variables, that are then used to constrain the expected behaviour in a more expressive way. In~\cite{DBLP:conf/atal/FerrandoAM17}, the same authors proposed an approach to verify AIPs at runtime using multiple monitors. This is obtained by decentralising the global specification (specified as a Trace Expression~\cite{DBLP:conf/birthday/AnconaFM16}), which is used to represent the global protocol, into partial specifications denoting the single agents' perspective. Both those approaches are based on the formalism serving as building block for RML's semantics. From this perspective, RV4JaCa is an evolution of these approaches in two ways: (i) it allows general-purpose verification since no constraint is assumed on the monitor side, except for being capable of receiving and sending JSON messages; (ii) to be self-contained, RV4JaCa natively supports RML monitors, and because of that it allows a more intuitive and high-level protocol specification.
In~\cite{bakar2013runtime,roungroongsom2015formal}, other approaches to runtime verification of agent interactions are proposed, and in~\cite{lim2016runtime} a framework for dynamic adaptive MAS (DAMS-RV) based on an adaptive feedback loop is presented.
Other approaches to MAS RV include the spin-off proposals from the SOCS project where the SCIFF computational logic framework~\cite{DBLP:conf/aiia/AlbertiGLMT05} is used to provide the semantics of social integrity constraints. To model MAS interaction, an expectation-based semantics specifies the links between observed and expected events, providing a means to test runtime conformance of an actual conversation with respect to a given interaction protocol~\cite{DBLP:books/igi/09/TorroniYS0CGLM09}. Similar work has been performed using commitments~\cite{Chesani:2009:CTV:1661445.1661461}.

To the best of our knowledge, RV4JaCa is the first approach that integrates RV within JaCaMo for verifying agent interaction protocols.

\section{Conclusions and Future Work}
\label{sec:conclusions}

Communications between agents play a key role in the functioning of a multi-agent system since, in practice, agents rarely act alone, they usually inhabit an environment that contains other agents. Therefore, an extra layer of security that allows us to verify key aspects of this message exchange adds great value, in addition to great possibilities for improvement, since certain aspects do not need to be considered when developing each of the agents. Using this type of formal verification at runtime allows us to standardise the interaction between agents through previously defined protocols that all agents must follow and, if they do not, react in a way that the execution is not negatively affected by the effects that were caused by this protocol deviation.

On the other hand, the checks done with RV are not limited to protocol validation. More specific properties of each application domain can also be verified once the monitor has access to the content of the exchanged messages. Even the execution of certain routines or functions according to the direction in which the conversations between the agents go can be done. For example, recording the results obtained during the agents' reasoning in a database without agents having the responsibility to carry out the registrations themselves. Or even sending an automatic email to a supervisor if any property identified by an agent and communicated to another is outside certain parameters.
Therefore, depending on the MAS's domain, there is a range of possibilities in which RV can be used.

Based on that, we proposed RV4JaCa, an approach to integrate multi-agent systems and runtime verification. Our approach was built using JaCaMo and RML, and it provides significant progress toward obtaining guarantees of the correct execution of the MAS. The case study presented in this paper demonstrates the use of RV4JaCa in practice, also showing promising preliminary results. In addition, it is important to note that our approach can be applied to different scenarios in different MAS.
Based on the presented case study, we created two distinct properties to be checked at runtime in the system. The first one is to allow agents to be alerted if there is an unexpected change in the topic of conversation by the human user who is interacting with the system. And the second one is capable of verifying that the previously defined communication protocol between agents is being followed correctly (in particular, question-answer relations).

As MAS have been used to build systems focused on explainability, RV's extra safety layer is certainly useful, since to achieve explainability in MAS much communication between agents and humans needs to be carried out. Also, we need to take into account that the developer does not have complete control over the interactions that human users can make during a dialogue. In this sense, RV allows us to avoid unexpected and probably inappropriate system behaviour.

As future work, we plan to further extend RV4JaCa to more than interaction protocols. For instance, it would be relevant to check the agents' state of mind as well. Moreover, through RV4JaCa we also want to create a library of verifiably correct agent interaction protocols to be used in different scenarios involving both agents and humans in the loop.

\bibliographystyle{eptcs}
\bibliography{generic}
\end{document}